\documentclass[11pt]{revtex4}

\usepackage[pdftex]{graphicx}
\usepackage{amsmath}
\usepackage{amsfonts}
\usepackage{amssymb}
\newcommand\spm{\mathrel{\text{\framebox[0.9\width]{$\pm$}}}}
\newcommand\smp{\mathrel{\text{\framebox[0.9\width]{$\mp$}}}}
\newcommand\cpm{\mathrel{\text{\textcircled{\makebox{$\pm$}}}}}
\newcommand\cmp{\mathrel{\text{\textcircled{\makebox{$\mp$}}}}}
\begin{document}
\title{Properties of families of spinors in $d=(5+1)$ with zweibein of an almost $S^2$ 
and two kinds of spin connection fields, allowing massless and massive solutions in 
$d=(3+1)$} 
\author{D. Lukman and N.S. Manko\v c Bor\v stnik \\
Department of Physics, FMF, University of Ljubljana,\\
Jadranska 19, SI-1000 Ljubljana, Slovenia
}

\begin{abstract} 
We studied in the refs.~\cite{hn,dn} properties of spinors in a toy model in $d=(5+1)$, 
when ${\cal M}^{(5+1)}$ breaks to an infinite disc with a zweibein which makes a disc 
curved on an almost $S^2$ and with a spin connection field which allows on such a sphere 
only one massless spinor state,  as a step towards realistic Kaluza-Klein  theories in 
non compact spaces. In the ref.\cite{dhn} we allow on $S^2$ two kinds of the spin 
connection fields, those which are gauge fields of spins in and those which
are the gauge fields of the family quantum numbers, both as required for this toy model 
by the {\it spin-charge-family} theory~\cite{norma92939495,JMP}. 
This time we study, by taking into account families of spinors interacting with 
several spin connection fields, properties of massless  and massive solutions of
equations of motion, with the discrete symmetries~\cite{HNds,TDN2013} 
($\mathbb{C}_{{ \cal N}}$, ${\cal P}_{{\cal N}}$, ${\cal T}_{{ \cal N}}$) included.
We also allow nonzero vacuum expectation values of the spin connection fields and study the 
masses.
\end{abstract}
\maketitle
\section{Introduction}
\label{introduction}
%
%

The {\it spin-charge-family} theory~\cite{norma92939495,JMP}, proposed by one of us (N.S.M.B.),   
is offering the explanation for the appearance of families of fermions in any dimension. Starting 
in $d=(13 + 1)$ with a simple action for massless fermions interacting with the gravitational 
interaction only - that is with the vielbeins and the two kinds of the spin connection fields, 
the ones originating in the Dirac kind of spin ($\gamma^a$'s) and the others originating in the 
second kind of the Clifford operators ($\tilde{\gamma}^a$'s) - the theory manifests effectively 
at low energies the observed properties of fermions and bosons, offering the explanation for all 
the assumptions of the {\it standard model}: For the appearance of families, for the appearance 
of the Higgs's scalar ~\cite{*Proc2014Scalars} with the weak and the hyper charges 
($\mp \frac{1}{2}$, $\pm \frac{1}{2}$, respectively), for the Yukawa couplings, for the charges 
of the family members, for the vector gauge fields, for the dark matter content, for the 
matter-antimatter asymmetry~\cite{*Proc2014Matter-antimatter}. 

The theory predicts the fourth family, which will soon be observed at the LHC, and several 
scalar fields, manifesting in the observed Higgs's scalar~\cite{*Proc2014fourfam} and the 
Yukawa couplings, some superposition of which will also be observed at the LHC.

A simple toy model~\cite{hn,dn,dhn}, which includes also families in the way proposed by the 
{\it spin-charge-family} theory~\cite{norma92939495,JMP}), is expected to help to better 
understand mechanisms causing the breaks of symmetries needed in the case of $d=(13+1)$, 
where a simple starting action leads in the low energy regime after the breaks to the 
observable phenomena. 
  
This contribution is a small further step in understanding properties of the families after 
the breaks of symmetries, caused by the scalar fields which are the gauge fields of the charges 
of spinors and the scalar fields which are the gauge fields of the family groups. The discrete 
symmetries of fermions and bosons in the case of only one family are studied already in  
the ref.~\cite{TDN2013}. Here the discrete symmetries are studied when the families are taken 
into account. We allow also that the spin connection fields gain nonzero vacuum expectation 
values and study solutions of the equations of motion for massive spinors.

We start with massless spinors~\cite{hn,dn,dhn,TDN2013} in  a flat manifold ${\cal M}^{(5+1)}$, 
which breaks into ${\cal M}^{(3+1)}$ times an infinite disc. The vielbein on the disc curves the 
disc into (almost) a sphere $S^{2}$
\begin{eqnarray}
e^{s}{}_{\sigma} = f^{-1}
\begin{pmatrix}1  & 0 \\
 0 & 1 
 \end{pmatrix},
f^{\sigma}{}_{s} = f
\begin{pmatrix}1 & 0 \\
0 & 1 \\
\end{pmatrix},
\label{fzwei}
\end{eqnarray}
with 
\begin{eqnarray}
\label{f}
f &=& 1+ (\frac{\rho}{2 \rho_0})^2= \frac{2}{1+\cos \vartheta},\nonumber\\ 
x^{(5)} &=& \rho \,\cos \phi,\quad  x^{(6)} = \rho \,\sin \phi, \quad E= f^{-2}.
\end{eqnarray}
The angle $\vartheta$ is the ordinary azimuthal angle on a sphere. 
The last relation follows  from $ds^2= 
e_{s \sigma}e^{s}{}_{\tau} dx^{\sigma} dx^{\tau}= f^{-2}(d\rho^{2} + \rho^2 d\phi^{2})$.
We use indices $(s,t)\in (5,6)$ to describe the flat index in the space of an infinite plane, and 
$(\sigma, \tau) \in ((5), (6)), $ to describe the Einstein index.  
Rotations around  the axis through the two poles of a sphere are described by the angle $\phi$, 
while $\rho = 2 \rho_0 \, \sqrt{\frac{1- \cos \vartheta}{1+ \cos \vartheta}}$. 
The volume of this non compact sphere is finite, equal to $V= \pi\, (2 \rho_0)^2$.  The symmetry 
of $S^2$ is a symmetry of $U(1)$ group. 

We take into account that there are two kinds of the Clifford algebra operators: Beside the Dirac 
$\gamma^a$ also $\tilde{\gamma}^a$, introduced in~\cite{norma92939495,JMP,holgernorma20023}. 
Correspondingly the covariant momentum of a spinor on an almost $S^{2}$ sphere is 
\begin{eqnarray}
\label{covmom}
p_{0a}&=& f^{\alpha}{}_a\, p_{\alpha} + \frac{1}{2E}\,\{p_{\alpha}, f^{\alpha}{}_a E\}_- 
-\frac{1}{2} S^{cd}  \omega_{cda}- \frac{1}{2} \tilde{S}^{cd} \tilde{\omega}_{cda}\,, \nonumber\\
S^{ab} &=& \frac{i}{4} (\gamma^a \gamma^b- \gamma^b \gamma^a)\,, \quad\quad  
\tilde{S}^{ab}= \frac{i}{4} (\tilde{\gamma}^a \tilde{\gamma}^b- \tilde{\gamma}^b \tilde{ \gamma}^a)\,,
\end{eqnarray}
with $ E = \det(e^a{\!}_{\alpha}) $ and with vielbeins $f^{\alpha}{\!}_{a}$~\footnote{$f^{\alpha}{}_{a}$ 
are inverted vielbeins to $e^{a}{}_{\alpha}$ with the properties $e^a{}_{\alpha} f^{\alpha}{\!}_b = $ 
$\delta^a{}_b,\; e^a{}_{\alpha} f^{\beta}{}_a = $ $\delta^{\beta}_{\alpha} $. 
Latin indices  
$a,b,..,m,n,..,s,t,..$ denote a tangent space (a flat index),
while Greek indices $\alpha, \beta,..,\mu, \nu,.. \sigma,\tau ..$ denote an Einstein 
index (a curved index). Letters  from the beginning of both the alphabets
indicate a general index ($a,b,c,..$   and $\alpha, \beta, \gamma,.. $ ), 
from the middle of both the alphabets   
the observed dimensions $0,1,2,3$ ($m,n,..$ and $\mu,\nu,..$), indices from 
the bottom of the alphabets
indicate the compactified dimensions ($s,t,..$ and $\sigma,\tau,..$). 
We assume the signature $\eta^{ab} =
diag\{1,-1,-1,\ldots,-1\}$.
}, the gauge fields of the infinitesimal generators of translation, and  with the two kinds of the 
spin connection fields: {\bf i.} $\omega_{ab\alpha}$,  the gauge fields of  $S^{ab}$ 
and {\bf ii.} $\tilde{\omega}_{ab\alpha}$,  the gauge fields of  $\tilde{S}^{ab}$. 

We make a choice  of the spin connection fields of the two kinds on the infinite disc as follows 
(assuming that there must be some fermion sources causing these spin connections, the study of 
such sources of the scalar fields $\omega_{st \sigma}$ and $\tilde{\omega}_{ab \sigma}$ are in 
progress)
\begin{eqnarray}
  f^{\sigma}{}_{s'}\, \omega_{st \sigma} &=& i F_{56}\, f \, \varepsilon_{st}\; 
  \frac{e_{s' \sigma} x^{\sigma}}{(\rho_0)^2}\,= 
  -\frac{1}{2E}\{p_{\sigma}, Ef^{\sigma}{}_{s'}\}_{-} \; \varepsilon_{st}\, 4 F_{56}\,, 
  \nonumber\\
  f^{\sigma}{}_{s'}\, \tilde{\omega}_{st \sigma} &=& i \tilde{F}_{56}\, f \, \varepsilon_{st}\; 
  \frac{e_{s' \sigma} x^{\sigma}}{(\rho_0)^2}\,
  = -\frac{1}{2E}\{p_{\sigma}, Ef^{\sigma}{}_{s'}\}_{-} \;\varepsilon_{st}\, 4 \tilde{F}_{56}\,, \nonumber\\
  f^{\sigma}{}_{s}\, \tilde{\omega}_{mn \sigma} &=&  -\frac{1}{2E}\{p_{\sigma}\,,  
  E f^{\sigma}{}_{s}\}_{-} \; 4 \tilde{F}_{mn}\;, \tilde{F}_{mn}= - \tilde{F}_{nm}\,,\nonumber\\
  &&\quad s=5,6,\,\,\; \sigma=(5),(6)\,. 
\label{omegas}
\end{eqnarray}
We take the starting action in agreement with the  {\it spin-charge-family} theory
for this toy model in  $d=(5+1)$, that is the action for a massless spinor  (${\cal S}_{f}$) with the 
covariant momentum $p_{0a}$ from Eq.~(\ref{covmom}) interacting with gravity only and for the 
vielbein and the two kinds of the spin connection  fields (${\cal S}_{b}$) 
  \begin{eqnarray}
         {\cal S}  = {\cal S}_{b} +  {\cal S}_{f}\,, 
  &\quad& {\cal S}_{f} = \int \; d^d x \, E \,{\cal L}_{f} \nonumber\\
  {\cal S}_{b} =  \int \; d^d x \, E \, (\alpha {\cal\,  R} + \tilde{\alpha} \tilde{{\cal\,  R}})\,, 
  &\quad &\,\;   {\cal L}_{f}= \psi^{\dagger}\, \gamma^0 \gamma^a \, p_{0a}\,\psi
  \,.
  \label{action}
  \end{eqnarray}
The two Riemann scalars,  ${\cal R} = {\cal R}_{abcd}\,\eta^{ac}\eta^{bd}$ and 
$\tilde{{\cal R}} = \tilde{{\cal R}}_{abcd}\,\eta^{ac}\eta^{bd}$, are
determined by the Riemann tensors
\begin{eqnarray}
     {\cal R}_{abcd}      &=& \frac{1}{2}\, f^{\alpha}{}_{[a} f^{\beta}{}_{b]}\;(\omega_{cd \beta, \alpha} 
- \omega_{ce \alpha} \omega^{e}{}_{d \beta} )\,, \nonumber\\ 
\tilde{{\cal R}}_{abcd}   &=& \frac{1}{2}\, f^{\alpha}{}_{[a} f^{\beta}{}_{b]}\;
(\tilde{\omega}_{cd \beta,\alpha} - \tilde{\omega}_{c e \alpha} \tilde{\omega}^{e}{}_{d \beta})\,,
\label{RtildeR}
\end{eqnarray}
where $[a\,\,b]$ means that the 
anti-symmetrization must be performed over the two indices $a$ and $b$.

We assume no gravity in $d=(3+1)$: $f^{\mu}{}_m = \delta^{\mu}_m$ and $\omega_{mn\mu}=0$ for 
$ m,n=(0,1,2,3), \;  \mu =(0,1,2,3)$. Accordingly $(a,b,\dots)$ run in Eq.~(\ref{action}) only over 
$s \in (5,6)$. 
Taking into account the subgroup structure of the operators  $\tilde{S}^{mn}$ 
\begin{eqnarray}
\label{so31sub}
\vec{\tilde{N}}_{(L,R)}= \vec{\tilde{N}}^{\cpm}&=&
\frac{1}{2} (\tilde{S}^{23}\pm i \tilde{S}^{01},\tilde{S}^{31}\pm i \tilde{S}^{02}, 
\tilde{S}^{12}\pm i \tilde{S}^{03} )\,, \nonumber\\
\tilde{N}_{L}{}^{\spm}\;\;=\tilde{N}^{\oplus \spm}&=&\tilde{N}^{\oplus 1}  \pm i\, \tilde{N}^{\oplus 2}\,, \quad 
\tilde{N}_{R}{}^{\spm}=\tilde{N}^{\ominus \spm}= \tilde{N}^{\ominus 1} \pm i\, \tilde{N}^{\ominus 2}\,,
\end{eqnarray}
we can rewrite the $\frac{1}{2} \tilde{S}^{cd} \tilde{\omega}_{cda}$ part of the covariant momentum 
(Eq.~\ref{covmom}) as follows 
\begin{eqnarray}
\label{omegamnFmn}
-\frac{1}{2}\, f \,  \tilde{S}^{mn} \tilde{\omega}_{mn\pm}  &=&
     \sum_i \tilde{N}^{\oplus i} \tilde{A}^{\oplus i}_{\pm} + \sum_i
     \tilde{N}^{\ominus i} \tilde{A}^{\ominus i}_{\pm} \nonumber\\
     &=&  \tilde{N}^{\oplus \boxplus } \tilde{A}^{\oplus \boxplus}_{\pm} + 
     \tilde{N}^{\oplus \boxminus } \tilde{A}^{\oplus \boxminus}_{\pm} +  
     \tilde{N}^{\oplus 3 } \tilde{A}^{\oplus 3}_{\pm}+ 
  \tilde{N}^{\ominus \boxplus } \tilde{A}^{\ominus \boxplus}_{\pm} + 
     \tilde{N}^{\ominus \boxminus } \tilde{A}^{\ominus \boxminus}_{\pm} +  
     \tilde{N}^{\ominus 3 } \tilde{A}^{\ominus 3}_{\pm
     } \,,\nonumber\\
     \tilde{\omega}_{mn\pm} &=& \tilde{\omega}_{mn 5} \mp i \tilde{\omega}_{mn 6}\,. 
\end{eqnarray}
The notation was used
\begin{eqnarray}
\label{so13}
\tilde{A}^{\cpm i}_{s}=f^{\sigma}{}_{s} \,\tilde{A}^{\cpm i}_{\sigma} &=&- f^{\sigma}{}_{s} \{
(\tilde{\omega}_{23 \sigma} \mp i \tilde{\omega}_{01 \sigma}), 
(\tilde{\omega}_{31 \sigma} \mp i \tilde{\omega}_{02 \sigma}),
(\tilde{\omega}_{12 \sigma} \mp i \tilde{\omega}_{03 \sigma})\}\,\nonumber\\
&=&  \, \delta^{\sigma}_{s }\; \frac{1}{2E}\{p_{\sigma}, Ef \}_{-} \; 4 \left ( 
\tilde{F}^{\cpm 1}, \tilde{F}^{\cpm 2}, \tilde{F}^{\cpm 3} \right)\,,\nonumber\\
\tilde{A}^{\oplus \spm}_{s} &=&  \frac{1}{2} \,(\tilde{A}^{\oplus 1}_{s}
\mp i \, \tilde{A}^{\oplus 2}_{s})\,,\quad 
\tilde{A}^{\ominus \spm}_{s} = \frac{1}{2} \,(\tilde{A}^{\ominus 1}_{s}
\mp i \, \tilde{A}^{\ominus 2}_{s})\,,\nonumber\\
\tilde{F}^{\oplus \spm}&=& (\tilde{F}^{23}\mp \tilde{F}^{02}) -
i (\pm \tilde{F}^{31}+\tilde{F}^{01})\,,\quad
\tilde{F}^{\oplus 3} = (\tilde{F}^{12}- i \tilde{F}^{03})\,, \nonumber\\
\tilde{F}^{\ominus \spm}&=& (\tilde{F}^{23}\pm \tilde{F}^{02}) + 
i (\mp \tilde{F}^{31}+\tilde{F}^{01})\,,\quad
\tilde{F}^{\ominus 3} = (\tilde{F}^{12}+ i \tilde{F}^{03})\,, \nonumber\\
\sigma &=& ((5),(6))\,,\quad s= (5,6) \,, 
\end{eqnarray}
with $\omega_{abc}$ and $\tilde{\omega}_{abc}$  defined in Eq.~(\ref{omegas}). 

We  looked in the ref.~\cite{dhn} for the  chiral fermions on this sphere, that is for the 
fermions of only one handedness in $d=(3+1)$ and accordingly mass protected, without including 
any extra fundamental gauge fields to the action from Eq.(\ref{action}). 

In this contribution we study  the influence of 
several spin connection fields on the properties of families,  
looking for the  intervals within   which the parameters  of both kinds of the spin connection 
fields ($F_{56}$, $\tilde{F}_{56}$, $\tilde{F}^{\oplus \spm} $,  $\tilde{F}^{\oplus 3}$, 
$\tilde{F}^{\ominus \spm} $, $\tilde{F}^{\ominus 3}$) allow massless solutions of the equation 
\begin{eqnarray}
&&\{\gamma^0 \gamma^m p_m +  f \gamma^0 \gamma^s \delta^{\sigma}_s  ( p_{0\sigma} 
+  \frac{1}{2 E f}
\{p_{\sigma}, E f\}_- )\}\psi=0,\quad {\rm with} \nonumber\\
&& p_{0\sigma} = p_{\sigma}- \frac{1}{2} S^{st}\omega_{st \sigma} - \frac{1}{2} \tilde{S}^{ab} 
\tilde{\omega}_{ab \sigma}\,, 
\label{weylE}
\end{eqnarray}
for  several families of spinors. 

We also allow nonzero vacuum expectation values of the scalar (with respect to $d=(3+1)$) gauge
fields and study properties of spinors. 

The discrete symmetries of the equations of motion and of solutions are studied in 
sections~(\ref{discrete}, \ref{DSofField}).

We look for the properties of spinors and gauge fields, scalars and vectors with respect to 
$d=(3+1)$.

In section~\ref{equations} we present spinor states in "our technique" (see appendix 
in the ref.~\cite{*Proc2014Matter-antimatter}). In section~\ref{masspm} we discuss massless and massive states of 
families of spinors. 
In section~\ref{discrete} we present discrete symmetry operators introduced in the refs.~\cite{HNds,TDN2013}, 
in section~\ref{DSofField} we discuss the properties of spinors and the gauge fields, the zweibein and 
the two kinds of the spin connection fields, under the discrete symmetry operators.

\section{Solutions of equations of motion  for families of spinors}
\label{equations}
%

We first briefly explain, following the refs.~\cite{JMP,hn,dn,dhn}, the appearance of families 
in our toy model, using what is called the technique~\cite{holgernorma20023}. 
%
 
There are $2^{d/2 -1}=4$ families in our toy model, each family with $2^{d/2 -1}=4$ members. 
In the technique~\cite{holgernorma20023} the states are defined as a product of nilpotents 
and projectors  
%
\begin{eqnarray}
\stackrel{ab}{(\pm i)}: &=& \frac{1}{2}(\gamma^a \mp  \gamma^b),  \; 
\stackrel{ab}{[\pm i]}: = \frac{1}{2}(1 \pm \gamma^a \gamma^b), \quad
{\rm for} \,\; \eta^{aa} \eta^{bb} = -1, \nonumber\\
\stackrel{ab}{(\pm )}: &= &\frac{1}{2}(\gamma^a \pm i \gamma^b),  \; 
\stackrel{ab}{[\pm ]}: = \frac{1}{2}(1 \pm i\gamma^a \gamma^b), \quad
{\rm for} \,\; \eta^{aa} \eta^{bb} =1,
\label{snmb:eigensab}
\end{eqnarray} 
which are the eigen vectors  of $S^{ab}$ as well as of $\tilde{S}^{ab}$ as follows
\begin{eqnarray}
S^{ab} \stackrel{ab}{(k)} =  \frac{k}{2} \stackrel{ab}{(k)}, \quad 
S^{ab} \stackrel{ab}{[k]} =  \frac{k}{2} \stackrel{ab}{[k]}, \quad
\tilde{S}^{ab} \stackrel{ab}{(k)}  = \frac{k}{2} \stackrel{ab}{(k)},  \quad 
\tilde{S}^{ab} \stackrel{ab}{[k]}  =   - \frac{k}{2} \stackrel{ab}{[k]}\;,
\label{snmb:eigensabev}
\end{eqnarray}
with the properties that $\gamma^a$ transform   
$\stackrel{ab}{(k)}$ into  $\stackrel{ab}{[-k]}$, while 
$\tilde{\gamma}^a$ transform  $\stackrel{ab}{(k)}$ 
into $\stackrel{ab}{[k]}$ 
\begin{eqnarray}
\gamma^a \stackrel{ab}{(k)}= \eta^{aa}\stackrel{ab}{[-k]},\; 
\gamma^b \stackrel{ab}{(k)}= -ik \stackrel{ab}{[-k]}, \; 
\gamma^a \stackrel{ab}{[k]}= \stackrel{ab}{(-k)},\; 
\gamma^b \stackrel{ab}{[k]}= -ik \eta^{aa} \stackrel{ab}{(-k)}\,,\nonumber\\
\label{snmbgraph}
\tilde{\gamma^a} \stackrel{ab}{(k)} = - i\eta^{aa}\stackrel{ab}{[k]},\;
\tilde{\gamma^b} \stackrel{ab}{(k)} =  - k \stackrel{ab}{[k]}, \;
\tilde{\gamma^a} \stackrel{ab}{[k]} =  \;\;i\stackrel{ab}{(k)},\; 
\tilde{\gamma^b} \stackrel{ab}{[k]} =  -k \eta^{aa} \stackrel{ab}{(k)}\,. 
\end{eqnarray}
After making a choice of the Cartan subalgebra, for which we take: ($S^{03}, S^{12}, S^{56}$) 
and ($\tilde{S}^{03}, \tilde{S}^{12}, \tilde{S}^{56}$), the four spinor families, each with 
four vectors, which are  eigen vectors of the chosen Cartan subalgebra with the eigen values 
from Eq.~(\ref{snmb:eigensabev})~\cite{dhn}, follow 
%
\begin{align}
\label{weylrep}
\varphi^{1 I}_{1} &= \stackrel{56}{(+)} \stackrel{03}{(+i)} \stackrel{12}{(+)}\psi_0,
&\varphi^{1 II}_{1} &= \stackrel{56}{(+)} \stackrel{03}{[+i]} \stackrel{12}{[+]}\psi_0, \nonumber\\
\varphi^{1 I}_{2} &=\stackrel{56}{(+)}  \stackrel{03}{[-i]} \stackrel{12}{[-]}\psi_0,
&\varphi^{1 II}_{2} &= \stackrel{56}{(+)}  \stackrel{03}{(-i)} \stackrel{12}{(-)}\psi_0,\nonumber\\
\varphi^{2 I}_{1} &=\stackrel{56}{[-]}  \stackrel{03}{[-i]} \stackrel{12}{(+)}\psi_0,
&\varphi^{2 II}_{1} &= \stackrel{56}{[-]}  \stackrel{03}{(-i)} \stackrel{12}{[+]}\psi_0,\nonumber\\
\varphi^{2 I}_{2} &=\stackrel{56}{[-]} \stackrel{03}{(+i)} \stackrel{12}{[-]}\psi_0,
&\varphi^{2 II}_{2} &= \stackrel{56}{[-]} \stackrel{03}{[+i]} \stackrel{12}{(-)}\psi_0,\nonumber\\
&&
\nonumber\\
\varphi^{1 III}_{1} &= \stackrel{56}{[+]} \stackrel{03}{[+i]}\stackrel{12}{(+)}\psi_0,
&\varphi^{1 IV}_{1} &= \stackrel{56}{[+]} \stackrel{03}{(+i)} \stackrel{12}{[+]}\psi_0,\nonumber\\
\varphi^{1 III}_{2} &=\stackrel{56}{[+]}  \stackrel{03}{(-i)}\stackrel{12}{[-]}\psi_0,
&\varphi^{1 IV}_{2} &=\stackrel{56}{[+]}  \stackrel{03}{[-i]} \stackrel{12}{(-)}\psi_0,\nonumber\\
\varphi^{2 III}_{1} &=\stackrel{56}{(-)}  \stackrel{03}{(-i)}\stackrel{12}{(+)}\psi_0,
&\varphi^{2 IV}_{1} &=\stackrel{56}{(-)}  \stackrel{03}{[-i]} \stackrel{12}{[+]}\psi_0,\nonumber\\
\varphi^{2 III}_{2} &=\stackrel{56}{(-)} \stackrel{03}{[+i]}\stackrel{12}{[-]}\psi_0,
&\varphi^{2 IV}_{2} &=\stackrel{56}{(-)}  \stackrel{03}{(+i)} \stackrel{12}{(-)}\psi_0,
\end{align}
where  $\psi_0$ is a vacuum  for the spinor state. One can reach  from the first member  
$\varphi^{1 I}_{1}$ of the first family the same  family member of  all the other families by 
the application of $\tilde{S}^{ab}$.
One can reach all the family members of each family 
by applying the generators $S^{ab}$ on one of the family  member. 
%
If we write the operators of handedness in $d=(5+1)$ as $\Gamma^{(5+1)} = \gamma^0 \gamma^1 
\gamma^2 \gamma^3 \gamma^5 \gamma^6$ ($= 2^3 i S^{03} S^{12} S^{56}$), in $d=(3+1)$ 
as $\Gamma^{(3+1)}= -i\gamma^0\gamma^1\gamma^2\gamma^3 $ ($= 2^2 i S^{03} S^{12}$) 
and in the two dimensional space as $\Gamma^{(2)} = i\gamma^5 \gamma^6$ 
($= 2 S^{56}$), we find that all the  states of all the families are left handed with respect to 
$\Gamma^{(5+1)}$, with the eigen value $-1$, the first two states of the first family, and 
correspondingly the first two states of any family, are right handed and the second two 
 states are left handed with respect to $\Gamma^{(2)}$, with  the eigen values $1$ and $-1$, 
 respectively, while the first two are left handed 
and the second two right handed with respect to $\Gamma^{(3+1)}$ with the eigen values $-1$ and $1$, 
respectively. 

Having the rotational symmetry around the axis perpendicular to the plane of the fifth and the sixth 
dimension we require that $\psi^{(6)}$ is the eigen function of the total angular momentum
operator $M^{56}= x^5 p^6-x^6 p^5  + S^{56}= -i \frac{\partial}{\partial \phi} + S^{56}$
\begin{eqnarray}
M^{56}\psi^{(6)}= (n+\frac{1}{2})\,\psi^{(6)}\,.
\label{mabx}
\end{eqnarray}

Accordingly we write, when taking into account Eq.~(\ref{weylrep}),  
the most general wave function  
$\psi^{(6)}$ obeying Eq.~(\ref{weylE}) in $d=(5+1)$ as
\begin{eqnarray}
\psi^{(6)}= {\cal N}\, \sum_{i=I,II,III,IV}\, ({\cal A}^{i}_{n}\, {\stackrel{56}{(+)}}{}^{i}\,
\psi^{(4i)}_{(+)}  
+ {\cal B}^{i}_{n+1}\, e^{i \phi}\, {\stackrel{56}{[-]}}{}^{i}\, \psi^{(4 i)}_{(-)})\, e^{in \phi}.
\label{mabpsi}
\end{eqnarray}
where ${\cal A}^{i}_{n}$ and ${\cal B}^{i}_{n}$ depend on $x^{\sigma}$, while $\psi^{(4 i)}_{(+)}$ 
and $\psi^{(4 i)}_{(-)}$  determine the spin 
and the coordinate dependent parts of the wave function $\psi^{(6)}$ in $d=(3+1)$ in accordance 
with the definition in Eq.(\ref{weylrep}), for example, 
\begin{eqnarray}
\psi^{(4 I)}_{(+)} &=& \alpha^{I}_+ \; {\stackrel{03}{(+i)}}\, {\stackrel{12}{(+)}} + 
\beta^{I}_+ \; {\stackrel{03}{[-i]}}\, {\stackrel{12}{[-]}}, \nonumber\\ 
\psi^{(4 I)}_{(-)}&=& \alpha^{I}_- \; {\stackrel{03}{[-i]}}\, {\stackrel{12}{(+)}} + 
\beta^{I}_- \; {\stackrel{03}{(+i)}}\, {\stackrel{12}{[-]}}. 
\label{psi4}
\end{eqnarray}
${\stackrel{56}{(+)}}{}^{i}= {\stackrel{56}{(+)}},$ for $i=I,II$ and ${\stackrel{56}{(+)}}{}^{i}=
{\stackrel{56}{[+]}}$ for $i=III,IV$, while ${\stackrel{56}{[-]}}{}^{i}= 
{\stackrel{56}{[-]}}$ for $i=I,II$ and ${\stackrel{56}{[-]}}{}^{i}= {\stackrel{56}{(-)}}$ for $i=III,IV$.
Using $\psi^{(6)}$ in Eq.~(\ref{weylE}) and separating dynamics in $(1+3)$ and on $S^2$, 
the following relations follow, from which we recognize the mass term $m^{I}$:  
$\,\frac{\alpha^{i}_{+}}{\alpha^{i}_-} (p^0-p^3) - \frac{\beta^{i}_+}{\alpha^{i}_-} (p^1-ip^2)= m^{i},\,$ 
$\,\frac{\beta^{i}_+}{\beta^{i}_-} (p^0+p^3) - \frac{\alpha^{i}_+}{\beta^{i}_-} (p^1+ip^2)= m^{i},\,$ 
$\,\frac{\alpha^{i}_-}{\alpha^{i}_+} (p^0+p^3) + \frac{\beta_-}{\alpha_+} (p^1-ip^2)= m^{i},\,$
$\,\frac{\beta^{i}_-}{\beta^{i}_+} (p^0-p^3) + \frac{\alpha^{i}_-}{\beta^{i}_+} (p^1-ip^2)= \,m^{i}.\,$ 
(One notices that for massless solutions  ($m^{i}=0$)  $\psi^{(4i)}_{(+)}$ 
and $\psi^{(4i)}_{(-)}$, for each $i=  I,II,III,IV,$ 
decouple.) 

For a spinor with the momentum $p^m= (p^0,0,0,p^3)$ in $d=(3+1)$ the spin and coordinate dependent parts 
for four families are:  $\psi^{(4 I)}_{(+)} = $ 
$\alpha \; {\stackrel{03}{(+i)}}\, {\stackrel{12}{(+)}}\,$, $\psi^{(4 II)}_{(+)} = $ 
$\alpha\; {\stackrel{03}{[+i]}}\, {\stackrel{12}{[+]}}\,$, $\psi^{(4 III)}_{(+)} = $ 
$\alpha\; {\stackrel{03}{[+i]}}\, {\stackrel{12}{(+)}}\,$, $\psi^{(4 IV)}_{(+)} = $ 
$\alpha \; {\stackrel{03}{(+i)}}\, {\stackrel{12}{[+]}}\,$.

Taking the above derivation into account (Eqs.~(\ref{mabpsi}, \ref{f}, \ref{omegas}, \ref{psi4},
\ref{so31sub}, \ref{omegamnFmn}, \ref{so13}))  the equation of motion for spinors follows~\cite{dhn} 
from the action~(\ref{action})
 \begin{eqnarray}
 \label{weylErho}
 if \,&& \{ e^{i \phi 2S^{56}}\, [(\frac{\partial}{\partial \rho} + \frac{i\, 2 S^{56}}{\rho} \, 
 (\frac{\partial}{\partial \phi}) ) -  \frac{1}{2 \,f} \, \frac{\partial f}{\partial \rho }\, 
 (1- 2 F_{56} \, 2S^{56} - 2 \tilde{F}_{56}\, 2\tilde{S}^{56}  \nonumber\\
 && -  2 \tilde{F}^{\ominus\,\boxplus}   \, 2 \tilde{N}^{\ominus\,\boxplus} -
       2 \tilde{F}^{\ominus\,\boxminus}  \, 2 \tilde{N}^{\ominus\,\boxminus}-
       2 \tilde{F}^{\ominus\,3}          \, 2 \tilde{N}^{\ominus\,3} 
       \nonumber\\
 && -  2 \tilde{F}^{\oplus\,\boxplus}    \, 2 \tilde{N}^{\oplus\,\boxplus} -
       2 \tilde{F}^{\oplus\,\boxminus}   \, 2 \tilde{N}^{\oplus\,\boxminus}-
       2 \tilde{F}^{\oplus\,3}           \, 2 \tilde{N}^{\oplus\,3}  )\,]
 \, \}  \psi^{(6)}\nonumber\\
 + && \gamma^0 \gamma^5 \, m \, \psi^{(6)}=0.
 \end{eqnarray}
One easily recognizes that, due to the break of ${\cal M}^{(5+1)}$ into ${\cal M}^{(3+1)} \times$ 
an infinite disc, which concerns  (by our assumption) both, $S^{ab}$ and $\tilde{S}^{ab}$ sector, there 
are two times two coupled families: The first and the second, and the third and the fourth, while 
the first and the second  remain decoupled from the third and the fourth.
We end up with  two decoupled groups of equations of motion~\cite{dhn} 
(which all depend on the parameters $F_{56}$ and $\tilde{F}_{56}$):\\
i. The equations for the first and the second family
\begin{eqnarray}
&&-if \bigl\{ [(\frac{\partial}{\partial\rho} - \frac{n}{\rho}) 
           - \frac{1}{2f} \frac{\partial f}{\partial\rho} 
           (1-2F_{56}-2\tilde{F}_{56}-2\tilde{F}^{\ominus 3})]\,  \mathcal{A}_n^I \nonumber\\
           &-& \frac{1}{2f} \frac{\partial f}{\partial\rho}
           \, 2\tilde{F}^{\ominus \boxplus}\, \mathcal{A}_n^{II}
    \bigr\} 
    + m \, \mathcal{B}_{n+1}^I = 0\,,\nonumber
\end{eqnarray}
\begin{eqnarray}
\label{mass12}
&&-if \bigl\{ [(\frac{\partial}{\partial\rho} + \frac{n+1}{\rho}) 
           - \frac{1}{2f} \frac{\partial f}{\partial\rho} 
                (1+2F_{56}-2\tilde{F}_{56} - 2\tilde{F}^{\ominus 3})] \, \mathcal{B}_{n+1}^{I} \nonumber\\
           &-& \frac{1}{2f} \frac{\partial f}{\partial\rho} 
           \, 2\tilde{F}^{\ominus \boxplus}\,  \mathcal{B}_{n+1}^{II}
   \bigr\} + m \, \mathcal{A}_{n}^{I} = 0\,,
\end{eqnarray}
\begin{eqnarray}
&&-if \bigl\{ [(\frac{\partial}{\partial\rho} - \frac{n}{\rho}) 
           - \frac{1}{2f} \frac{\partial f}{\partial\rho} 
            (1-2F_{56}-2\tilde{F}_{56}+ 2\tilde{F}^{\ominus 3})] \, \mathcal{A}_n^{II} \nonumber\\
           &-& \frac{1}{2f} \frac{\partial f}{\partial\rho}
           \, 2\tilde{F}^{\ominus \boxminus}\, \mathcal{A}_n^{I}
    \bigr\} + m \,\mathcal{B}_{n+1}^{II} = 0\,,\nonumber
    \end{eqnarray}
    \begin{eqnarray}
&&-if \bigl\{ [(\frac{\partial}{\partial\rho} + \frac{n+1}{\rho}) 
           - \frac{1}{2f} \frac{\partial f}{\partial\rho} 
              (1+2F_{56}-2\tilde{F}_{56}+2\tilde{F}^{\ominus 3})] \, \mathcal{B}_{n+1}^{II} \nonumber\\
           &-& \frac{1}{2f} \frac{\partial f}{\partial\rho} 
              \, 2\tilde{F}^{\ominus \boxminus}\, \mathcal{B}_{n+1}^{I}
    \bigl\} + m \,\mathcal{A}_{n}^{II} = 0\,.\nonumber
\end{eqnarray}
ii. The equations for the third and the fourth family 
\begin{eqnarray}
&&-if \bigl\{ [(\frac{\partial}{\partial\rho} - \frac{n}{\rho}) 
         - \frac{1}{2f} \frac{\partial f}{\partial\rho}
           (1-2F_{56}+2\tilde{F}_{56} -2\tilde{F}^{\oplus 3})] \, \mathcal{A}_n^{III}\nonumber\\
         &-& \frac{1}{2f} \frac{\partial f}{\partial\rho}
          \, (-2\tilde{F}^{\oplus \boxplus})\, \mathcal{A}_n^{IV}
    \bigr\} + m \,\mathcal{B}_{n+1}^{III} = 0\,,\nonumber
    \end{eqnarray}
    \begin{eqnarray}
\label{mass34}
&&-if \bigl\{ [(\frac{\partial}{\partial\rho} + \frac{n+1}{\rho}) 
          - \frac{1}{2f} \frac{\partial f}{\partial\rho} 
               (1+2F_{56}+2\tilde{F}_{56}-2\tilde{F}^{\oplus 3})] \, \mathcal{B}_{n+1}^{III}\nonumber\\
           &-& \frac{1}{2f} \frac{\partial f}{\partial\rho} 
             \,(-2\tilde{F}^{\oplus \boxplus })\, \mathcal{B}_{n+1}^{IV}
    \bigl\} + m \, \mathcal{A}_{n}^{III} = 0\,,
    \end{eqnarray}
\begin{eqnarray}
&&-if \bigl\{ [(\frac{\partial}{\partial\rho} - \frac{n}{\rho}) 
           - \frac{1}{2f} \frac{\partial f}{\partial\rho} 
           (1-2F_{56}+2\tilde{F}_{56}+2\tilde{F}^{\oplus 3})] \, \mathcal{A}_n^{IV}\nonumber\\
           &-& \frac{1}{2f} \frac{\partial f}{\partial\rho} 
             \,(-2\tilde{F}^{\oplus \boxminus })\, \mathcal{A}_n^{III}
    \bigr\} + m \,\mathcal{B}_{n+1}^{IV} = 0\,,\nonumber
    \end{eqnarray}
    \begin{eqnarray}
%
&&-if \bigl\{ [(\frac{\partial}{\partial\rho} + \frac{n+1}{\rho}) 
           - \frac{1}{2f} \frac{\partial f}{\partial\rho} 
                  (1+2F_{56}+2\tilde{F}_{56}+2\tilde{F}^{\oplus 3})] \, \mathcal{B}_{n+1}^{IV}\nonumber\\
           &-& \frac{1}{2f} \frac{\partial f}{\partial\rho} 
            \,(-2\tilde{F}^{\oplus \boxminus})\, \mathcal{B}_{n+1}^{III}
    \bigr\} + m \,\mathcal{A}_{n}^{IV} = 0\,. \nonumber
\end{eqnarray}
Let us look for possible normalizable~\cite{hn,dn} massless solutions for each of the two groups 
in dependence on the parameters which determine the strength of the spin connection fields. Both 
groups, although depending on different parameters of the spin connection fields, can be treated in 
an equivalent way. Let us therefore study massless solutions of the first group of equations of motion.

For $m=0$   the equations for $\mathcal{A}_n^{I}$ and $\mathcal{A}_n^{II}$ in Eq.~(\ref{mass12}) 
decouple from those for $\mathcal{B}_{n+1}^{I}$ and $\mathcal{B}_{n+1}^{II}$.
We get for massless solutions
\begin{eqnarray}
\label{group12Amassless}
\mathcal{A}_{n}^{I\pm }\;\,&=& a_{\pm}\;
\rho^{n}\, f^{\frac{1}{2}(1- 2F_{56} -2\tilde{F}_{56})}\;
f^{ \pm \sqrt{(\tilde{F}^{\ominus 3})^2+\tilde{F}^{\ominus \boxplus} \tilde{F}^{\ominus \boxminus}}} \,,\nonumber\\ 
\mathcal{A}_{n}^{II\pm}\;&=&  
\frac{\pm \sqrt{(\tilde{F}^{\ominus 3})^2+\tilde{F}^{\ominus \boxplus} \tilde{F}^{\ominus \boxminus}}
+ \tilde{F}^{\ominus 3}}{\tilde{F}^{\ominus \boxplus}}\;\mathcal{A}_{n}^{I\pm }\,,
\nonumber\\
\mathcal{B}_{n+1}^{I \pm}&=& b_{\pm}\,
\rho^{-n-1}\, f^{\frac{1}{2}(1 + 2F_{56} -2\tilde{F}_{56})}\,
f^{\pm \sqrt{(\tilde{F}^{\ominus 3})^2+
\tilde{F}^{\ominus \boxplus} \tilde{F}^{\ominus \boxminus}}} \,,\nonumber\\ 
\mathcal{B}_{n+1}^{II\pm}&=& 
\frac{\pm \sqrt{(\tilde{F}^{\ominus 3})^2+\tilde{F}^{\ominus \boxplus} \tilde{F}^{\ominus \boxminus}}
+ \tilde{F}^{\ominus 3}}{\tilde{F}^{\ominus \boxplus}}\;\mathcal{B}_{n+1}^{I\pm }\,,
\end{eqnarray}
$n$ is a positive integer.  The solutions ($\mathcal{A}_{n}^{I + }$, $\mathcal{A}_{n}^{II + }$) 
and ($\mathcal{A}_{n}^{I - }$, $\mathcal{A}_{n}^{II - }$) are two independent solutions, a general solution
is any superposition of these two. Similarly is true for 
($\mathcal{B}_{n+1}^{I \pm}$, $\mathcal{B}_{n+1}^{II \pm}$). 

In the massless case also $\mathcal{A}_{n}^{I,II\pm} $ 
decouple from $\mathcal{B}_{n+1}^{I,II\pm} $.

One can easily write down massless solutions of the second group of two families, 
decoupled from the first one, when knowing massless solutions of the first group of families. It follows
\begin{eqnarray}
\label{group12Bmassless}
\mathcal{A}_{n}^{III \pm}\;\,&=& a_{\pm}\;
\rho^{n}\, f^{\frac{1}{2}(1- 2F_{56} + 2\tilde{F}_{56})}\;
f^{ \pm \sqrt{(\tilde{F}^{\oplus 3})^2+\tilde{F}^{\oplus \boxplus} \tilde{F}^{\oplus \boxminus}}} \,,\nonumber\\ 
\mathcal{A}_{n}^{IV \pm}\;&=&  
\frac{\pm \sqrt{(\tilde{F}^{\oplus 3})^2+\tilde{F}^{\oplus \boxplus} \tilde{F}^{\oplus \boxminus}}
+ \tilde{F}^{\oplus 3}}{-\tilde{F}^{\oplus \boxplus}}\;\mathcal{A}_{n}^{III \pm}\,,
\nonumber\\
\mathcal{B}_{n+1}^{III \pm}&=& b_{\pm}\,
\rho^{-n-1}\, f^{\frac{1}{2}(1 + 2F_{56} +2\tilde{F}_{56})}\,
f^{\pm \sqrt{(\tilde{F}^{\oplus 3})^2+
\tilde{F}^{\oplus \boxplus} \tilde{F}^{\oplus \boxminus}}} \,,\nonumber\\ 
\mathcal{B}_{n+1}^{IV \pm}&=& 
\frac{\pm \sqrt{(\tilde{F}^{\oplus 3})^2+\tilde{F}^{\oplus \boxplus} \tilde{F}^{\oplus \boxminus}}
+ \tilde{F}^{\oplus 3}}{-\tilde{F}^{\oplus \boxplus}}\;\mathcal{B}_{n+1}^{III\pm }\,,
\end{eqnarray}
$n$ is a positive integer, $ a_{\pm}$ and $ b_{\pm}$ are normalization factors.

Requiring that only normalizable (square integrable) solutions are acceptable 
\begin{eqnarray}
2\pi \, \int^{\infty}_{0} \,E\, \rho d\rho \,({\cal A}^{i \star}_{n} {\cal A}^{i}_{n}
+{\cal B}^{i\star}_{n} {\cal B}^{i}_{n} )&& < \infty\,, 
\label{masslesseqsolf}
\end{eqnarray}
$i\in\{I,II,III,IV\}\,$, one finds that $\mathcal{A}^{i}_{n}$ and $\mathcal{B}^{i}_{n}$ are 
normalizable~\cite{hn,dn} under the following conditions 
\begin{eqnarray}
\label{norm-solAB1234}
& & \mathcal{A}^{I,II}_{n}\;\;\;\;\,: \;\,-1 < n < 2 (F_{56}+ \tilde{F}_{56} 
\pm \sqrt{(\tilde{F}^{\ominus 3})^2+ \tilde{F}^{\ominus \boxplus} \tilde{F}^{\ominus \boxminus} }\,\,)\,, 
\nonumber\\
& &\mathcal{B}^{I,II}_{n}\;\;\;\;\,:\;\, 2(F_{56} - \tilde{F}_{56}\pm  \sqrt{(\tilde{F}^{\ominus 3})^2+
 \tilde{F}^{\ominus \boxplus} \tilde{F}^{\ominus \boxminus} }\,\,)< n <1 \, ,
 \nonumber\\
& &\mathcal{A}^{III,IV}_{n}\,: \;\,-1 < n < 2 (F_{56}- \tilde{F}_{56} 
\pm \sqrt{(\tilde{F}^{\oplus 3})^2+ \tilde{F}^{\oplus \boxplus} \tilde{F}^{\oplus \boxminus} }\,\,)\,, 
\nonumber\\
& &\mathcal{B}^{III,IV}_{n}\,:\;\, 2(F_{56} + \tilde{F}_{56}\pm  \sqrt{(\tilde{F}^{\oplus 3})^2+
 \tilde{F}^{\oplus \boxplus} \tilde{F}^{\oplus \boxminus} }\,\,)< n <1 \, .
\end{eqnarray}
One immediately sees that for $F_{56}=0=\tilde{F}_{56}$ there is no solution for the zweibein from Eq.~(\ref{f}). 
Let us first assume that $\tilde{F}^{\cmp i}=0\,; \,i\in\{1,2,3\}$.  
Eq.~(\ref{norm-solAB1234}) tells us that the strengths $F_{56}, \tilde{F}_{56}$ of the spin connection fields 
($\omega_{56 \sigma}$ and $\tilde{\omega}_{56 \sigma}$) can make a choice between the 
massless solutions (${\cal A}^{I,II}_n$, ${\cal A}^{III,IV}_n$) and (${\cal B}^{I,II}_n,  {\cal B}^{III,IV}_n$):\\ 
For 
\begin{eqnarray}
0< 2(F_{56}+ \tilde{F}_{56}) \le 1,\quad \tilde{F}_{56} < F_{56}
\label{Fformassless}
\end{eqnarray}
there exist four massless left handed solutions  with respect 
to $(3+1)$. For  
\begin{eqnarray}
0< 2(F_{56}+ \tilde{F}_{56}) \le 1,\quad \tilde{F}_{56} = F_{56}
\label{Fformassless2}
\end{eqnarray}
 the only massless solution are the two left handed spinors with respect to 
$(3+1)$
\begin{eqnarray}
\psi^{(6 \;I,II)m=0}_{\frac{1}{2}} ={\cal N}_0  \; f^{-F_{56}\,- \tilde{F}_{56}+1/2} 
\stackrel{56}{(+)}\psi^{(4 \; I,II)}_{(+)}.
\label{Massless}
\end{eqnarray} 
The solutions~(Eq.\ref{Massless}) are the eigen functions  of $M^{56}$ with the eigen value $1/2$. 
Since no right handed massless solutions are allowed, the left handed ones are mass protected. 
For the  particular choice  $2(F_{56} + \tilde{F}_{56})=1$ the spin connection fields  
$-S^{56} \omega_{56\sigma} - \tilde{S}^{56} \tilde{\omega}_{56\sigma}$ 
compensate the term $\frac{1}{2Ef} \{p_{\sigma}, Ef \}_- $ and the  left handed spinor
with respect to $d=(1+3)$ becomes a constant with respect to $\rho $ and $\phi$.  
To make one of these two states massive, one can try to include terms like 
$\tilde{F}^{\cmp i}$.

Let us keep $\tilde{F}^{\oplus i}=0\; i\in\{1,2,3\}$ and $F_{56}= \tilde{F}_{56}$, while we take 
$\tilde{F}^{\ominus 3}\,,\tilde{F}^{\ominus \smp}$ non zero.
Now it is still true that due to the conditions in Eq.~(\ref{norm-solAB1234})  there are no 
massless solutions determined by ${\cal A}^{III,IV}$ and ${\cal B}^{III,IV}$. 
There is now only one massless and mass protected family for $F_{56}= \tilde{F}_{56}$. In this case the 
solutions $\mathcal{A}^{I -}_{0} $ and $\mathcal{A}^{II -}_{0} $ are related
\begin{eqnarray}
\label{masslessAm}
\mathcal{A}^{I -}_{0} &=& {\cal N}^{-
}_{0} \, f^{\frac{1}{2}[1-2F_{56}-2\tilde{F}_{56}-
      2 \sqrt{(\tilde{F}^{\ominus 3})^2+\tilde{F}^{\ominus \boxplus} \tilde{F}^{\ominus \boxminus}}]} 
      \,,\nonumber\\ 
\mathcal{A}^{II -}_{0} &=& - \frac{(\sqrt{(\tilde{F}^{\ominus 3})^2+ \tilde{F}^{\ominus \boxplus}
\tilde{F}^{\ominus \boxminus} }+ \tilde{F}^{\ominus 3})}{\tilde{F}^{\ominus \boxplus}}\,\mathcal{A}^{I -}_{0}\,.
\end{eqnarray}
There exists, however, one additional massless state, with $\mathcal{A}^{I +}_{0} $  
related to $\mathcal{A}^{II + }_{0} $ and $\mathcal{B}^{I +}_{0} $ 
related to $\mathcal{B}^{II + }_{0} $, which fulfil Eq.~(\ref{norm-solAB1234}). But since we have left 
and right handed massless solution present, it is not mass protected any longer.

One can make  a choice as well that none of solutions would be massless. 

According to Eq.~(\ref{CemptNPNcheck3}) from sect.~\ref{DSofField} the equation of motion presented in 
Eqs.~(\ref{action}, \ref{covmom}) are covariant with respect to the discrete symmetry operator
$\mathbb{C}_{{\cal N}} \cdot {\cal P}_{{\cal N}}$ (Eq.~(\ref{CemptNPN})), what means that the 
antiparticle feels the transformed gauge fields and carry the opposite charge with respect to the 
starting particle.

Let us conclude this section by recognizing that for $\tilde{F}^{\oplus \spm}=0 $ and  
$\tilde{F}^{\ominus \spm} =0$ all the families decouple. There is then the choice of the parameters 
($F_{56}\,$, $\tilde{F}_{56}\,$, $\tilde{F}^{\oplus 3}$, $\tilde{F}^{\ominus 3}\, $) which determine 
how many massless and mass protected families exist, if  any.

\subsection{Solutions after the scalar gauge fields gain nonzero vacuum expectation values}
\label{masspm}

Let us now assume that the spin connection fields gain nonzero vacuum expectation values
\begin{eqnarray}
\label{wvactildew}
m^{(56)}_{\pm}:&=& <\omega_{56\pm}>\,,\quad \tilde{m}^{(56)}_{\pm}:= <\tilde{\omega}_{56\pm}>\,,\nonumber\\
\tilde{m}^{(\tilde{N}_{R}i)}_{\pm}:&=& (<\tilde{\omega}_{23\pm} - i \tilde{\omega}_{01\pm}>,
<\tilde{\omega}_{31\pm} - i \tilde{\omega}_{02\pm}>,<\tilde{\omega}_{12\pm} - i \tilde{\omega}_{03\pm}>)\,,\nonumber\\
\tilde{m}^{(\tilde{N}_{L}i)}_{\pm}:&=& (<\tilde{\omega}_{23\pm} + i \tilde{\omega}_{01\pm}>,
<\tilde{\omega}_{31\pm} + i \tilde{\omega}_{02\pm}>,<\tilde{\omega}_{12\pm} + i \tilde{\omega}_{03\pm}>)\,,
\end{eqnarray}
breaking the charge $S^{56}$ symmetry, as well as all the "tilde charges" ($\tilde{S}^{56}$, 
$\vec{\tilde{N}}^{(R,L)}$).

Then the equation of motion (\ref{weylE}) can be rewritten as
\begin{eqnarray}
\label{m+-}
& &\{ \gamma^0 \gamma^m p_{0m} + \gamma^0 \, \sum_{+,-} \stackrel{56}{(\pm)}\,
p_{0\pm}\}\,\psi=0\,,\nonumber\\
& &p_{0m}= p_m - S^{56}\, \omega_{56m}\,,\nonumber\\
& &p_{0\pm}= p_{\pm} - S^{56}\, m^{(56)}_{\pm} - \tilde{S}^{56} \,\tilde{m}^{(56)}_{\pm} 
              - \sum^{3}_{i=1}\tilde{N}^{i}_{R}\, \tilde{m}^{(\tilde{N}_{R} i)}_{\pm}
              - \sum^{3}_{i}\tilde{N}^{i}_{L}\, \tilde{m}^{(\tilde{N}_{R} i)}_{\pm}\,.
\end{eqnarray}
One finds that requiring the hermiticity of the equations of motion (Eq.~(\ref{m+-})) leads to the
relations
%
\begin{eqnarray}
\label{mher}
-m^{(56)}_{+} &=& m^{(56)}_{-}\,, \quad \tilde{m}^{(56)}_{+}=\tilde{m}^{(56)}_{-}\,,\quad
\quad \tilde{m}^{(\tilde{N}_{R}i)}_{+}=\tilde{m}^{(\tilde{N}_{R}i)}_{-}\,,\quad
\tilde{m}^{(\tilde{N}_{L}i)}_{+}=\tilde{m}^{(\tilde{N}_{L}i)}_{-}\,.
\end{eqnarray}
We also must require, to be consistent with the definition and the Eqs.~(\ref{CemptNPN}, 
\ref{CemptNPNcheck1}, \ref{CemptNPNcheck2}, \ref{CemptNPNcheck3}) and Eq.~(\ref{m+-}), that 
%
\begin{eqnarray}
\label{mCP}
 \mathbb{C}_{{\cal N}} {\cal P}_{{\cal N}}\, m^{(56)}_{\pm}\, 
(\mathbb{C}_{{\cal N}} {\cal P}_{{\cal N}})^{-1} &=&- m^{(56)}_{\mp}\,, \nonumber\\
 \mathbb{C}_{{\cal N}} {\cal P}_{{\cal N}}\, \tilde{m}^{(56)}_{\pm}\, 
(\mathbb{C}_{{\cal N}} {\cal P}_{{\cal N}})^{-1} &=& \tilde{m}^{(56)}_{\mp}\,, \nonumber\\
 \mathbb{C}_{{\cal N}} {\cal P}_{{\cal N}}\, \tilde{m}^{(\tilde{N}_{R}i)}_{\pm}\, 
(\mathbb{C}_{{\cal N}} {\cal P}_{{\cal N}})^{-1} &=& \tilde{m}^{(\tilde{N}_{R}i)}_{\mp}\,, 
\nonumber\\
 \mathbb{C}_{{\cal N}} \cdot {\cal P}_{{\cal N}}\, \tilde{m}^{(\tilde{N}_{L}i)}_{\pm}\, 
(\mathbb{C}_{{\cal N}} \cdot {\cal P}_{{\cal N}})^{-1} &=& \tilde{m}^{(\tilde{N}_{L}i)}_{\mp}\,.
\end{eqnarray}

Eq.~(\ref{m+-}) has then the solutions
\begin{eqnarray}
\label{msol}
m^{\tilde{N}_{L}}_{1,2}&= &\frac{1}{2}\,(m^{(56)}_{-}- \tilde{m}^{(56)}_{-})\pm 
\sqrt{\sum_{i}(\tilde{m}^{(\tilde{N}_{L}i)}_{-})^2}\,,\nonumber\\
m^{\tilde{N}_{R}}_{1,2}&= &\frac{1}{2}\,(m^{(56)}_{-}+ \tilde{m}^{(56)}_{-})\pm 
\sqrt{\sum_{i}(\tilde{m}^{(\tilde{N}_{R}i)}_{-})^2}\,,
\end{eqnarray}
with the spinor states with no conserved charge $S^{56}$ any longer
\begin{eqnarray}
\label{psisol}
\psi^{(6)\tilde{N}_{L}}_{m(1,2)}&= & N^{\tilde{N}_{L}}\,\,
\{ (\tilde{m}^{(\tilde{N}_{L}3)}_{-}\; \pm \;\sqrt{\,\sum_{i} \,(\tilde{m}^{(\tilde{N}_{L}i)}_{-}\,)^2\;} \,)\,
(\,\stackrel{03}{[+i]} \stackrel{12}{(+)} \stackrel{56}{[+]} - \stackrel{03}{(-i)} \stackrel{12}{(+)} \stackrel{56}{(-)})
\nonumber\\
&+& (\tilde{m}^{(\tilde{N}_{L}1)}_{-} + i \tilde{m}^{(\tilde{N}_{L}2)}_{-} )\,
(\,\stackrel{03}{(+i)} \stackrel{12}{[+]} \stackrel{56}{[+]} - \stackrel{03}{[-i]} \stackrel{12}{[+]} \stackrel{56}{(-)})\}
e^{-imx^0}\,,\nonumber\\
\psi^{(6)\tilde{N}_{R}}_{m(1,2)}&= & N^{\tilde{N}_{R}}\,\,
\{ (\tilde{m}^{(\tilde{N}_{R}3)}_{-}\; \mp \;\sqrt{\sum_{i} \, (\tilde{m}^{(\tilde{N}_{R}i)}_{-}\,)^2\;}\, )\,
(\,\stackrel{03}{[+i]} \stackrel{12}{[+]} \stackrel{56}{(+)} - \stackrel{03}{(-i)} \stackrel{12}{[+]} \stackrel{56}{[-]})
\nonumber\\
&+& (\tilde{m}^{(\tilde{N}_{R}1)}_{-} -i \tilde{m}^{(\tilde{N}_{R}2)}_{-} )\,
(\,\stackrel{03}{(+i)} \stackrel{12}{(+)} \stackrel{56}{(+)} - \stackrel{03}{[-i]} \stackrel{12}{(+)} \stackrel{56}{[-]})\}
e^{-imx^0}\,,
\end{eqnarray}
while handedness in the "tilde" sector is conserved.
%


%
\section{Discrete symmetries of spinors and gauge fields of the toy model}
\label{DSofField}

In the subsection of this  section~\ref{discrete} the discrete symmetry operators for 
particles and antiparticles  in the second quantized  picture are presented, as well as 
for the gauge fields. This definition for the discrete symmetry operators, as they  
manifest from the point of view of $d=(3+1)$, is designed for all the Kaluza-Klein 
like theories. At least this way of looking for the appropriate discrete symmetry operators 
from the point of view of $d=(3+1)$ can be helpful in all the Kaluza-Klein cases to find the 
appropriate discrete symmetry operators in the observable dimensions.

One sees that the operators of discrete symmetries, presented in Eqs.~(\ref{CPTN}, 
\ref{empt}, \ref{CemptPTN}), do not depend on the family quantum numbers $\tilde{\gamma}^{a}$, 
which means that every particle, described as a member of one family, transforms under the 
product of the two discrete symmetry operators $\mathbb{C}_{{ \cal N}}$ and 
$ {\cal P}_{{ \cal N}}$, presented in Eq.~(\ref{CPTN}) and Eq.~(\ref{CemptPTN}), into the 
corresponding antiparticle state, which belongs to the same family (carrying the same 
family quantum numbers). 

The discrete symmetry operator $\mathbb{C}_{{ \cal N}}$ $\cdot {\cal P}_{{ \cal N}}$ (Eqs.~(\ref{CPTN},
\ref{CemptPTN})) is in our case with $d=(5+1)$ equal to 
\begin{eqnarray}
\label{CemptNPN}
\mathbb{C}_{{\cal N}} \cdot {\cal P}_{{\cal N}}&  = & \gamma^0\,\gamma^{5}\,I_{\vec{x}_{3}}\,
I_{x_{6}}\,.
\end{eqnarray}
It has an even number of $\gamma^a$'s, which guarantees that the operation does not cause the 
transformation into another Weyl representation in $d=(5+1)$, which means that we stay within the 
Weyl representation from which we started.

Let us check what does this discrete symmetry operator 
$\mathbb{C}_{{ \cal N}}$ $\cdot {\cal P}_{{ \cal N}}$ do when being applied 
on several operators.

One easily finds
\begin{eqnarray}
\label{CemptNPNcheck1}
\mathbb{C}_{{ \cal N}}\cdot {\cal P}_{{ \cal N}}\, (\gamma^0,\gamma^1,\gamma^2,\gamma^3,
\gamma^5,\gamma^6)\,(\mathbb{C}_{{ \cal N}}\cdot {\cal P}_{{ \cal N}})^{-1}
& = & (-\gamma^0,\gamma^1,\gamma^2,\gamma^3,-\gamma^5,\gamma^{6})\,,\nonumber\\
\mathbb{C}_{{ \cal N}}\cdot {\cal P}_{{ \cal N}}\, (p^0,p^1,p^2,p^3,p^5,p^6)\,
(\mathbb{C}_{{ \cal N}}\cdot {\cal P}_{{ \cal N}})^{-1}
& = & (p^0,-p^1,-p^2,-p^3,p^5,-p^{6})\,,\nonumber\\
\mathbb{C}_{{ \cal N}}\cdot {\cal P}_{{ \cal N}}\, \stackrel{56}{(\pm)} 
(\mathbb{C}_{{ \cal N}}\cdot {\cal P}_{{ \cal N}})^{-1}
& = & \stackrel{56}{(\mp)} \,,\nonumber\\
\mathbb{C}_{{ \cal N}}\cdot {\cal P}_{{ \cal N}}\, \tilde{\gamma^a}\,
(\mathbb{C}_{{ \cal N}}\cdot {\cal P}_{{ \cal N}})^{-1}
& = & \tilde{\gamma}^a \,,\; {\rm for}\, \,{\rm each} \,\,a\,,\nonumber\\
\mathbb{C}_{{ \cal N}}\cdot {\cal P}_{{ \cal N}}\, (\omega_{565}(x^0, \vec{x}_{3},x^5,x^6),
\omega_{566}(x^0, \vec{x}_{3},x^5,x^6))\,
(\mathbb{C}_{{ \cal N}}\cdot {\cal P}_{{ \cal N}})^{-1}
& = & \nonumber\\
(-\omega_{565}(x^0, -\vec{x}_{3},x^5,-x^6) ,\omega_{566} (x^0, -\vec{x}_{3},x^5,-x^6))\,,&& \nonumber\\
\mathbb{C}_{{ \cal N}}\cdot {\cal P}_{{ \cal N}}\, (\tilde{\omega}_{\tilde{5}\tilde{6}5} (x^0, \vec{x}_{3},x^5,x^6),
\tilde{\omega}_{\tilde{5}\tilde{6}6}(x^0 ,\vec{x}_{3},x^5,x^6))\,
(\mathbb{C}_{{ \cal N}}\cdot {\cal P}_{{ \cal N}})^{-1}
& = & \nonumber\\
(\tilde{\omega}_{\tilde{5}\tilde{6}5} (x^0, - \vec{x}_{3},x^5,-x^6),
-\tilde{\omega}_{\tilde{5}\tilde{6}6} (x^0, -\vec{x}_{3},x^5,-x^6))\,,&& 
\end{eqnarray}
where we write $\tilde{\omega}_{\tilde{5}\tilde{6} s}, s=(5,6)$ to point out that the first two 
indices  belong to the $\widetilde{SO}(5,1)$ group. We also use the notation $\stackrel{56}{(\pm)} =$
$\frac{1}{2}\,(\gamma^5 \pm i \gamma^{6}) $. 

One correspondingly finds, taking into account Eqs.~(\ref{so31sub}, \ref{so13})
\begin{eqnarray}
\label{CemptNPNcheck2}
\mathbb{C}_{{ \cal N}}{\cal P}_{{ \cal N}}\,\;S^{56} \,\;
(\mathbb{C}_{{ \cal N}}{\cal P}_{{ \cal N}})^{-1}
& = & - S^{56} \,,\nonumber\\
\mathbb{C}_{{ \cal N}}{\cal P}_{{ \cal N}}\,\;\tilde{S}^{56} \,\;
(\mathbb{C}_{{ \cal N}} {\cal P}_{{ \cal N}})^{-1}
& = &  \tilde{S}^{56} \,,\nonumber\\
\mathbb{C}_{{ \cal N}}{\cal P}_{{ \cal N}}\,\;\omega_{56m} (x^0,\vec{x}_{3}) \,\;
(\mathbb{C}_{{ \cal N}}\cdot {\cal P}_{{ \cal N}})^{-1}
& = & - \omega_{56m} (x^0,-\vec{x}_{3}) \,,\nonumber\\
\mathbb{C}_{{ \cal N}}{\cal P}_{{ \cal N}}\,\;(\vec{\tilde{A}}^{\cpm}_{5} (x^5,x^6),
\vec{\tilde{A}}^{\cpm}_{6} (x^5,x^6))\,\;
(\mathbb{C}_{{ \cal N}}{\cal P}_{{ \cal N}})^{-1}
& = & (\vec{\tilde{A}}^{\cpm}_{5} (x^5,-x^6),
-\vec{\tilde{A}}^{\cpm}_{6} (x^5,-x^6)) \,,\nonumber\\
\mathbb{C}_{{ \cal N}} {\cal P}_{{ \cal N}}\,\;\vec{\tilde{A}}^{\cpm}_{\pm} (x^5,x^6)\,\;
(\mathbb{C}_{{ \cal N}} {\cal P}_{{ \cal N}})^{-1}
& = & \vec{\tilde{A}}^{\cpm}_{\mp} (x^5,-x^6) \,,\nonumber\\
\mathbb{C}_{{ \cal N}} {\cal P}_{{ \cal N}}\,\;A^{56}_{\pm} (x^5,x^6)\,\;
(\mathbb{C}_{{ \cal N}}{\cal P}_{{ \cal N}})^{-1}
& = & - A^{56}_{\mp} (x^5,-x^6) \,,
\end{eqnarray}
with $A^{56}_{\pm}=$ $(\omega_{565}\mp i\omega_{566})$.

From  Eqs.~(\ref{CemptNPNcheck1}, \ref{CemptNPNcheck2}) it follows
\begin{eqnarray}
\label{CemptNPNcheck3}
&& \mathbb{C}_{{ \cal N}}\cdot {\cal P}_{{ \cal N}}\,\;\{ \gamma^0 \gamma^m \,(p_{m} - 
S^{56}\, \omega_{56m}(x^0,\vec{x}_{3}))\}\,\;
(\mathbb{C}_{{ \cal N}}\cdot {\cal P}_{{ \cal N}})^{-1}\nonumber\\
 &&=  \{ (-\gamma^0) (-\gamma_m )\,\;(p^{m} - 
(-S^{56})\,(- \omega_{56}{}^{m})(x^0,-\vec{x}_{3}))\} = \{ \gamma^0 \gamma^m \,(p_{m} - 
S^{56}\, \omega_{56m} (x^0,-\vec{x}_{3}))\}\,\nonumber\\
&&\mathbb{C}_{{ \cal N}}\cdot {\cal P}_{{ \cal N}}\,\;\{ \gamma^0 \stackrel{56}{(\pm)}\,(p_{\pm}  
-S^{56}\, \omega_{56\pm}(x^5, x^6)- \frac{1}{2}\,\tilde{S}^{\tilde{a}\tilde{b}} \,
\tilde{\omega}_{\tilde{a} \tilde{b} \pm} (x^5, x^6))\}\,\;
(\mathbb{C}_{{ \cal N}}\cdot {\cal P}_{{ \cal N}})^{-1}\nonumber\\
 &&=  \{ \gamma^0 \stackrel{56}{(\mp)}\,(p_{\mp}  
- S^{56}\, \omega_{56\mp}(x^5,- x^6)- \frac{1}{2}\,\tilde{S}^{ab}\, \tilde{\omega}_{\tilde{a} \tilde{b}\mp} 
(x^5,- x^6))\}\,.
\end{eqnarray}
Taking into account Eq.~(\ref{omegas}) and the equations of~(\ref{CemptNPNcheck3}),  
we see that the equations of motion are covariant with respect to a particle and its 
antiparticle: A particle and 
its antiparticle carry the same mass, while the antiparticle carries the opposite charge $S^{56}$ than 
the particle and moves in the transformed $U(1)$ field
$-\omega_{56}{}^{m} (x^0,-\vec{x}^{3})$~\cite{ItZu}. 

The equations of motion for our toy model (Eqs.~(\ref{weylE},\ref{omegas}), and 
correspondingly the solutions (Eq.~(\ref{mabpsi})) manifest the discrete symmetries 
$\mathbb{C}_{{ \cal N}} \cdot $ $ {\cal P}_{{ \cal N}} $, ${\cal T}_{{ \cal N}}$ and 
$\mathbb{C}_{{ \cal N}} \cdot $ ${\cal P}_{{ \cal N}} $ $\cdot {\cal T}_{{ \cal N}}$, 
with the operators presented in 
Eqs.~(\ref{CPTN}, \ref{CemptPTN}). Both, ${\cal C}_{{ \cal N}} \cdot {\cal P}_{{ \cal N}} $
$ \cdot \Psi^{(6)}$  and $\mathbb{C}_{{ \cal N}}$ $\cdot {\cal P}_{{ \cal N}} \Psi^{(6)}$  
(\ref{CemptPTN}) solve the equations of motion, provided that 
$\omega_{56m}(x^0,\vec{x}_{3})$ is a real field. 
The field $\omega_{56m}(x^0,\vec{x}_{3})$ transforms under ${\cal C}_{{ \cal N}}$ 
$ \cdot {\cal P}_{{ \cal N}} $ and $\mathbb{C}_{{ \cal N}} $ $\cdot {\cal P}_{{\cal N}}$ to 
$-\omega_{56}{}^{m}(x^0,-\vec{x}_{3})$, like the $U(1)$ field must~\cite{ItZu}. 


The starting action~(\ref{action}) and the corresponding Weyl equation~(\ref{weylE}) 
manifest discrete symmetries $\mathbb{C}_{{ \cal N}} \cdot $ 
$ {\cal P}_{{ \cal N}} $, ${\cal T}_{{ \cal N}}$ and $\mathbb{C}_{{ \cal N}} \cdot $
${\cal P}_{{ \cal N}} $ $\cdot {\cal T}_{{ \cal N}}$ from Eqs.~(\ref{CPTN},\ref{CemptPTN}). 
Correspondingly all the states with the conserved charges $M^{56}$ respect this symmetry, 
transforming particle states into the antiparticle states.

\subsection{Discrete symmetry operators} 
\label{discrete}

To discuss properties of the representations of particle and antiparticle states and of the gauge 
fields with which spinors interact let us first define the discrete symmetry operators  as seen from
the point of view of $d=(3+1)$ in the second quantized picture as proposed in the ref.~\cite{HNds}, 
where the definition of the discrete symmetries operators for the Kaluza-Klein kind of theories, for 
the first and the second quantized picture was defined, so that the total angular moments in higher 
dimensions manifest as charges in $d=(3+1)$. The ref.~\cite{HNds} uses the Dirac sea second quantized 
picture to make presentation transparent.

The ref.~\cite{HNds} proposes the following discrete symmetry operators
\begin{eqnarray}
\label{CPTN}
{\cal C}_{{\cal N}}  &= & \prod_{\Im \gamma^m, m=0}^{3} \gamma^m\,\, \,\Gamma^{(3+1)} \,
K \,I_{x^6,x^8,\dots,x^{d}}  \,,\nonumber\\
{\cal T}_{{\cal N}}  &= & \prod_{\Re \gamma^m, m=1}^{3} \gamma^m \,\,\,\Gamma^{(3+1)}\,K \,
I_{x^0}\,I_{x^5,x^7,\dots,x^{d-1}}\,,\nonumber\\
{\cal P}^{(d-1)}_{{\cal N}}  &= & \gamma^0\,\Gamma^{(3+1)}\, \Gamma^{(d)}\, I_{\vec{x}_{3}}
\,.
\end{eqnarray}
The operator of handedness in even $d$ dimensional spaces is defined as
\begin{eqnarray}
\label{handedness}
\Gamma^{(d)} :=(i)^{d/2}\; \prod_a \:(\sqrt{\eta^{aa}}\, \gamma^a)\,,
\end{eqnarray}
 with products of 
$\gamma^a$ in ascending order. We choose $\gamma^0$, $\gamma^1$ real, $\gamma^2$ imaginary, 
$\gamma^3$ real, $\gamma^5$ imaginary, $\gamma^6$ real, alternating imaginary and real up to 
$\gamma^d$ real. 
Operators $I$ operate as follows: $\quad I_{x^0} x^0 = -x^0\,$; $
I_{x} x^a =- x^a\,$; $  I_{x^0} x^a = (-x^0,\vec{x})\,$; $ I_{\vec{x}} \vec{x} = -\vec{x}\,$; $
I_{\vec{x}_{3}} x^a = (x^0, -x^1,-x^2,-x^3,x^5, x^6,\dots, x^d)\,$; 
$I_{x^5,x^7,\dots,x^{d-1}}$ $(x^0,x^1,x^2,x^3,x^5,x^6,x^7,x^8,
\dots,x^{d-1},x^d)$ $=(x^0,x^1,x^2,x^3,-x^5,x^6,-x^7,\dots,-x^{d-1},x^d)$; $I_{x^6,x^8,\dots,x^d}$ 
$(x^0,x^1,x^2,x^3,x^5,x^6,x^7,x^8,\dots,x^{d-1},x^d)$
$=(x^0,x^1,x^2,x^3,x^5,-x^6,x^7,-x^8,\dots,x^{d-1},-x^d)$, $d=2n$. 

${\cal C}_{{\cal N}}$ transforms the state, put on the top of the Dirac sea, into the corresponding 
negative energy state in the Dirac sea.

We need the operator, we name~\cite{NBled2013,TDN2013,HNds} it $\mathbb{C}_{{ \cal N}}$, which transforms 
the starting single particle state on the top of the Dirac sea into the negative energy state and 
then empties this negative energy state.  
This hole in the Dirac sea is the antiparticle state  
put on the top of the Dirac sea. Both, a particle and its antiparticle state (both put on the top of 
the Dirac sea), must solve the Weyl equations of motion.

This $\mathbb{C}_{{ \cal N}}$ is defined as a product of the operator~\cite{NBled2013,TDN2013} $"emptying"$,
(which is really an useful operator, although it is somewhat difficult to imagine it, since it is making transformations 
into a completely different Fock space)
\begin{eqnarray}
\label{empt}
"emptying"&=& \prod_{\Re \gamma^a}\, \gamma^a \,K =(-)^{\frac{d}{2}+1} \prod_{\Im \gamma^a}\gamma^a \,
\Gamma^{(d)} K\,, 
\end{eqnarray}
and ${\cal C}_{{\cal N}}$
\begin{eqnarray}
\label{CemptPTN}
\mathbb{C}_{{ \cal N}} &=& 
\prod_{\Re \gamma^a, a=0}^{d} \gamma^a \,\,\,K
\, \prod_{\Im \gamma^m, m=0}^{3} \gamma^m \,\,\,\Gamma^{(3+1)} \,K \,I_{x^6,x^8,\dots,x^{d}}\nonumber\\
&=& 
\prod_{\Re \gamma^s, s=5}^{d} \gamma^s \, \,I_{x^6,x^8,\dots,x^{d}}\,.
\end{eqnarray}

Let us present also the second quantized notation, following the notation in the ref.~\cite{HNds}.
Let  ${\underline {\bf {\Huge \Psi}}}^{\dagger}_{p}[\Psi_{p}]$ be the creation  operator  creating  
a fermion in the state $\Psi_{p}$  and let ${\mathbf{\Psi}}^{\dagger}_{p}(\vec{x})$
be the second quantized field creating a fermion at position $\vec{x}$. Then 
\begin{eqnarray}
\{ {\underline {\bf {\Huge  \Psi}}}^{\dagger}_{p}[\Psi_{p}] &=& \int \, 
{\mathbf{\Psi}}^{\dagger}_{p}(\vec{x})\, 
\Psi_{p}(\vec{x}) d^{(d-1)} x \,\} 
\,|vac> \nonumber
\end{eqnarray}
so that the antiparticle  state becomes
\begin{eqnarray}
\{ {\underline {\bf \mathbb{C}}}_{{\bf {\cal N}}}\, 
{\underline {\bf {\Huge \Psi}}}^{\dagger}_{p}[\Psi_{p}] &=&  
\int \, {\mathbf{\Psi}}_{p}(\vec{x})\, 
({\cal C}_{{\cal N}} \,\Psi_{p} (\vec{x})) d^{(d-1)} x \} \, |vac> \,.\nonumber
\end{eqnarray}
The antiparticle operator 
${\underline {\bf {\Huge \Psi}}}^{\dagger}_{a}[\Psi_{p}]$,  to the corresponding  particle  
creation operator, can also be written as 
\begin{eqnarray}
\label{makingantip}
{\underline {\bf {\Huge \Psi}}}^{\dagger}_{a}[\Psi_{p}]\, |vac>  &=& 
{\underline {\bf \mathbb{C}}}_{{{\bf \cal N}}}\, 
{\underline {\bf {\Huge \Psi}}}^{\dagger}_{p}[\Psi_{p}]\, |vac>  =  
\int \, {\mathbf{\Psi}}^{\dagger}_{a}(\vec{x})\, 
({\bf \mathbb{C}}_{\cal N}\,\Psi_{p} (\vec{x})) \,d^{(d-1)} x  \, \,|vac> \,,\nonumber\\
{\bf \mathbb{C}}_{\cal H} &=& "emptying"\,\cdot\, {\cal C}_{{\cal H}}  \,.
\end{eqnarray}

While the discrete symmetry operator $\mathbb{C}_{{ \cal N}}$  has an odd number of 
$\gamma^{a}$ operators and correspondingly transforms one Weyl representation in $d=(5+1)$ 
into another Weyl representation in $d=(5+1)$, changing the handedness of the representation,
stays the operator $\mathbb{C}_{{ \cal N}}$ $\cdot {\cal P}_{{ \cal N}}$  within the same 
Weyl. The same is true for  ${\cal T}_{{ \cal N}}$  and also for the product 
$\mathbb{C}_{{ \cal N}}$ $\cdot {\cal P}_{{ \cal N}}\cdot$ ${\cal T}_{{ \cal N}}$.

\section{Conclusions and discussions}
\label{conclusion}

We make in this contribution a small step further with respect to the refs.~\cite{dhn,TDN2013} in  
understanding the existence of massless and mass protected spinors as well as the massive states 
in non compact spaces 
in the presence of families of spinors after breaking symmetries. 
We take  a toy model in ${\cal M}^{5+1}$, which breaks into ${\cal M}^{3+1} \times$  an 
infinite disc curled into an almost $S^2$ under the influence of the zweibein. Following the 
{\it spin-charge-family} theory we have in this toy model four families. We study properties of 
families when allowing that besides  
the spin connection field, which are the gauge field of $S^{st}=\frac{i}{4}(\gamma^s \gamma^t-
\gamma^t \gamma^s)$,  also the gauge fields of $\tilde{S}^{st}=\frac{i}{4}
(\tilde{\gamma}^s \tilde{\gamma}^t- \tilde{\gamma}^t \tilde{\gamma}^s)$, determining families,   
affect the behaviour of spinors.

We simplify our study by assuming the same radial dependence of all the spin connection 
fields (Eq.~(\ref{omegas})), while the strengths of the fields ($F_{56}$, $\tilde{F}_{56}$,
$\tilde{F}_{mn}$) are allowed to vary within some intervals. 

We found that the choices of the parameters allow within some intervals of parameters ($F_{56}$, 
$\tilde{F}_{56}$, $\tilde{F}_{mn}$) four, two or none massless  and mass protected spinors. 

We allowed the nonzero vacuum expectation values of all the spin connection fields,
$f^{\sigma}{}_{s'}$ $\omega_{56 \sigma}$, $f^{\sigma}{}_{s'}$ $\tilde{\omega}_{\tilde{5}\tilde{6} \sigma}$ 
and $f^{\sigma}{}_{s'}$ $\tilde{\omega}_{\tilde{m}\tilde{n} \sigma}$, where $\sigma= ((5),(6)),$ $s=(5,6)$,
$\tilde{m}=(\tilde{0},\tilde{1},\tilde{2},\tilde{3})$. All indices $\tilde{a}$ belong to the 
$\widetilde{SO}(5,1)$ group, while indices $a$ belong to the $SO(5,1)$ group. The nonzero vacuum expectation
values of all the gauge fields causes that the $U(1)$ charge ($S^{56}$) breaks, as well as also all the 
family quantum numbers, while the handedness in the "tilde" degrees of freedom  keep two groups 
of families non coupled. 

We studied also the discrete symmetries of equations of motion and of solutions, for massless and 
massive states.

We found:
{\bf a.} $\;\;$
Almost $S^2$ or any other shape with the symmetry around the axis, 
perpendicular to the infinite disc, has the rotational symmetry around 
this axis. But almost $S^2$ has not the rotational symmetry around the axis 
which goes through the centre of almost sphere because of the singular point 
on the southern pole unless we make the translation of the axis. 
Equivalently the almost torus - infinite disc curled into an almost torus - 
has no symmetry. 
{\bf b.} $\;\;$ Even number of families stay massless  and mass protected for the 
intervals of parameters.  
{\bf c.} $\;\;$ Non zero vacuum expectation  values of the scalar gauge fields break all 
the charges, while the two handedness in the "tilde" sector keeps the two groups of families 
separated.
{\bf d.} $\;\;$ Let us add that while the weak charge and the hyper charge have fractional 
values in the {\it spin-cahreg-family} theory in $d=(13+1)$, have the scalar fields in this 
case of $d=(5+1)$ integer valued charges.


\begin{thebibliography}{99}
\bibitem{hn} D. Lukman, N. S. Manko\v c Bor\v stnik, H. B. Nielsen, "An effective two 
               dimensionality" cases bring a new hope to the Kaluza-Klein-like theories'', 
               http://arxiv.org/abs/1001.4679v5,
               {\it New J. Phys.} {\bf 13} (2011) 103027. arXiv:1001.4679v4.
\bibitem{dn} D. Lukman, N. S. Manko\v c Bor\v stnik, "Spinor states on a curved infinite disc 
                   with non-zero spin-connection fields", http://arxiv.org/abs/1205.1714, {\it J. of Phys.A: 
               Math. Theor. }{\bf 45} (2012) 465401 (19pp), doi:10.1088/1751-8113/45/46/465401.
\bibitem{dhn} D. Lukman, N. S. Manko\v c Bor\v stnik, H. B. Nielsen, "Families of Spinors in $d=(1+5)$ 
               with Zweibein and Two Kinds of Spin Connection Fields   on an almost $S^2$",  
               Proceedings to the 
                  $15^{th}$ international workshop ''What Comes Beyond 
		                the Standard Models'', July 9-19, 2012, Ed. N.S. Manko\v c Bor\v stnik, 
                H.B. Nielsen, D. Lukman, DMFA  Zalo\v zni\v stvo, Ljubljana December 2012, 
                 p. 163,  [arXiv:1012.0224, arXiv:1212.2370].  
\bibitem{norma92939495} N. S. Manko\v c Bor\v stnik, 
Phys. Lett. {\bf B 292} (1992) 25,
%
 J. Math. Phys. {\bf 34} (1993) 3731,
%
 Int. J. Theor. Phys. {\bf 40} 315 (2001),
%
Modern Phys. Lett.  {\bf A 10} (1995) 587, 
%
A. Bor\v stnik, N. S. Manko\v c Bor\v stnik, 
hep-ph/0401043, hep-ph/0401055, hep-ph/0301029,
%
Phys. Rev.  { \bf D 74} (2006)073013, hep-ph/0512062, 
%
G. Bregar, M. Breskvar, D. Lukman, N.S. Manko\v c Bor\v stnik,
		    hep-ph/0711.4681.
\bibitem{JMP}	N.S. Manko\v c Bor\v stnik, 
{\it J. of Modern Phys.} {\bf 4},   823 (2013) [arxiv:1312.1542], 
		    %
                     New J. of Phys. {\bf 10} (2008) 093002,  
                     hep-ph/0606159, hep/ph-07082846, 
                      hep-ph/0612250, p.25-50.
\bibitem{*Proc2014Scalars} N.S. Manko\v c Bor\v stnik,"The {spin-charge-family} theory explains 
why the scalar Higgs carries the weak  charge $\mp \frac{1}{2}$ and the hyper charge $\pm \frac{1}{2}$",
in this Proceedings, [http://arxiv.org/abs/1409.4981].
\bibitem{*Proc2014Matter-antimatter} N.S. Manko\v c Bor\v stnik, "Can spin-charge-family theory explain baryon number 
              non conservation?", in this Proceedings [arxiv:1409.7791].
\bibitem{*Proc2014fourfam} G. Bregar, N.S. Manko\v c Bor\v stnik, "The new experimental data for the quarks mixing 
               matrix are in better agreement with the {\it spin-charge-family} theory predictions", in this 
               Proceedings [arxiv:]
\bibitem{HNds} N.S. Manko\v c Bor\v stnik, H.B. Nielsen, "Discrete 
                   symmetries in the Kaluza-Klein-like theories", JHEP 04 (2014) 165
                   http://arxiv.org/abs/1212.2362v2. 
 \bibitem{TDN2013} T. Troha, D. Lukman, N.S. Manko\v c Bor\v stnik, 
              {\it Int. J. of Mod. Phys.} {\bf A 29} 1450124 (2014),
    arxiv:1312.1541. 
%
\bibitem{NBled2013}  N.S. Manko\v c Bor\v stnik, 
Spin-charge-family theory is explaining appearance of families of quarks and leptons, of Higgs 
and Yukawa couplings, in {\it Proceedings  to the 16th Workshop "What comes beyond the 
standard models", Bled, 14-21 of July, 2013}, eds. N.S. Manko\v c Bor\v stnik, H.B. Nielsen 
 and D. Lukman (DMFA  Zalo\v zni\v stvo, Ljubljana, December 2013) p.113 
[arxiv:1312.1542].
%
%
\bibitem{holgernorma20023} N. S. Manko\v c Bor\v stnik, H. B. Nielsen,
                          {\it J. of Math. Phys.} {\bf 43} (2002) 5782, [hep-th/0111257], 
                          {\it J. of Math. Phys.} {\bf 44} (2003) 4817, [hep-th/0303224].
                          

%
\bibitem{kk} T. Kaluza, Sitzungsber. Preuss. Akad. Wiss. Berlin, Math. Phys. {\bf 96} (1921) 69,
O. Klein, Z.Phys. {\bf 37} (1926) 895.
%
%
\bibitem{hnkk06} N. S. Manko\v c Bor\v stnik, H. B. Nielsen, 
Phys. Lett. {\bf B 633} (2006) 771-775, hep-th/0311037, hep-th/0509101,
%
 Phys. Lett. {\bf B 644} (2007)198-202, 
                     hep-th/0608006, 
%
                      Phys.  Lett. {\bf B 10} (2008)1016,
%
N. S. Manko\v c Bor\v stnik, H. B. Nielsen, D. Lukman, 
			hep-ph/0412208.
%
\bibitem{ItZu} C. Itzykson and J. Zuber, {\it Quantum Field Theory} (McGraw-Hill 1980). 
%
			%
\end{thebibliography}
\end{document}